\documentclass[final,5p,times,twocolumn]{elsarticle}

\usepackage[utf8]{inputenc}
\usepackage[T1]{fontenc}
\usepackage{xcolor}

\usepackage{amssymb}
\usepackage{amsmath,amsfonts}%
\usepackage{booktabs}
\usepackage{multirow}
\usepackage{hyperref}

\usepackage[normalem]{ulem}
\usepackage{enumitem}

\newcommand{\abs}[1]{\lvert #1\rvert}

\journal{Sensors and Actuators B: Chemical}

\begin{document}

\begin{frontmatter}

\title{High-Accuracy Material Classification via Reference-Free Terahertz Spectroscopy: Revisiting Spectral Referencing and Feature Selection}

\author[1]{Mathias Hedegaard Kristensen\corref{cor1}}
\cortext[cor1]{Corresponding author. Presently at CROMA UMR 5130 (CNRS, Grenoble INP, Univ. Grenoble Alpes, Univ. Savoie Mont Blanc), 73370 Le Bourget du Lac, France.}
\ead{mhkr@icloud.com}
\affiliation[1]{organization={Department of Materials and Production, Aalborg University},
            city={Aalborg East},
            postcode={DK-9220}, 
            country={Denmark}}
\author[1]{Paweł Piotr Cielecki}
\author[1]{Esben Skovsen}

\begin{abstract}
We investigate how feature selection algorithms can enable accurate, reference-free classification of materials using sparse-frequency terahertz (THz) reflection spectroscopy. 
Three classes of feature selection strategies are evaluated. Namely, the filter-based mRMR (minimum Redundancy Maximum Relevance), the embedded LASSO (Least Absolute Shrinkage and Selection Operator), and the wrapper-based SFS (Sequential Forward Selection) algorithms.
Each strategy is assessed using the Linear Logistic Regression, Naïve Bayes, and Support Vector Machine classifiers.
Our results show that high classification accuracy can be achieved using only a small subset of frequencies. Particularly, when non-referenced spectra are applied.
Furthermore, we show that the SFS-selected features align with the materials' absorption bands, confirming that the discriminative power arises from genuine spectroscopic contrasts. 
These findings highlights that reducing spectral dimensionality through data-driven selection eliminates the need for broadband sources and reference measurements, enabling compact, application-specific THz sensors.
This approach offers robust material identification in real-world scenarios such as security screening, non-destructive testing, and environmental monitoring.
\end{abstract}


\end{frontmatter}


\section{Introduction}
Terahertz (THz) spectroscopy offers a unique, non-invasive means to identify materials based on their characteristic spectral fingerprints, with promising applications in security screening, industrial quality control, and environmental monitoring \cite{Jepsen11,Naftaly19}.
THz radiation’s ability to penetrate non-metallic and non-polar materials makes reflection-mode measurements particularly attractive for practical sensing scenarios where transmission is not feasible.
However, THz reflection measurements yield weak and broad spectroscopic fingerprints due to their dependence on the refractive index \cite{Palka2011}.
In addition, system-related artifacts and atmospheric water vapor absorption further complicates the spectra, necessitating reference measurements for correction.
This dependence on referencing limits sensor deployment in real-world, dynamic environments where acquiring accurate reference spectra is challenging or impossible.
To address this limitation, only a few studies have explored reference-free approaches. 
Techniques such as phase imaging in transmission\cite{Zhang2008} and reflection geometries\cite{Zhong2008} have been proposed, but these methods still suffer from sensitivity to environmental factors and have not fully eliminated the need for references under realistic conditions.

Machine learning (ML) techniques, including Bayesian classifiers,\cite{Cao20,Nowak19} neural networks,\cite{Liu18,Zhong06,Zhang20} support vector machines,\cite{Wang17,Liu16,Knyazkova20} and random forests\cite{Cao20,Liu18,Liu16}, have shown promise for THz spectral classification. 
Recently, Park \emph{et al.} \cite{Park2021} reviewed the application of ML techniques in THz time-domain spectroscopy (TDS) and THz imaging, highlighting the pivotal role of ML in advancing THz technologies.
Yet, these studies have relied on referenced spectra, limiting their applicability in real-world settings.

Dimensionality reduction is critical for efficient ML-based THz sensing due to the high spectral dimensionality (hundreds to thousands of frequency components). While feature extraction methods like Principal Component Analysis (PCA) and Linear Discriminant Analysis (LDA) reduce dimensionality by transforming data, feature selection techniques identify a subset of the original frequencies most relevant for classification \cite{Aggarwal2014,Alpaydin14}. 
This distinction is crucial for sensor design, as selecting a reduced set of highly discriminative frequencies enables accurate material classification using fewer data points, improving algorithmic efficiency and directly informing hardware requirements.
In particular, it facilitates the development of narrowband THz sensors that operate only at the most informative frequencies. 
Such sensors can employ compact, high-power electronic THz sources in place of broadband systems, thereby improving signal-to-noise ratio and robustness while reducing cost and complexity.

Building on our earlier work evaluating linear feature extraction methods such as PCA and LDA for classifying THz reflection spectra \cite{Cielecki2021,Kristensen2022_IRMMWTHz}, and our recent demonstration of high-accuracy (>98.6\%) reference-free classification using Regularized LDA \cite{Kristensen2024}, this study systematically investigates feature selection strategies – filter-based, embedded, and wrapper-based – for reference-free THz reflection spectroscopy. 
We benchmark three algorithms, viz. minimum Redundancy Maximum Relevance (mRMR), Least Absolute Shrinkage and Selection Operator (LASSO), and Sequential Forward Selection (SFS), in combination with commonly used classifiers: Logistic Regression (LR), Naïve Bayes (NB), and Support Vector Machine (SVM).

Our goals are twofold: (1) to evaluate the accuracy and robustness of feature selection algorithms on raw, non-referenced spectra compared to referenced spectra; and (2) to identify the minimal number of THz frequencies needed for reliable material classification. By achieving high-accuracy classification with only a few selected frequencies, this approach paves the way toward compact, efficient THz sensors better suited for real-world applications where broadband sources and reference measurements are unavailable or impractical.

The implications of this work for sparse-frequency THz sensing hardware will be explored in future studies.

\section{Materials and Methods}
\subsection{Experimental Setup and Materials}
\begin{figure}
\center
\includegraphics[width=\columnwidth]{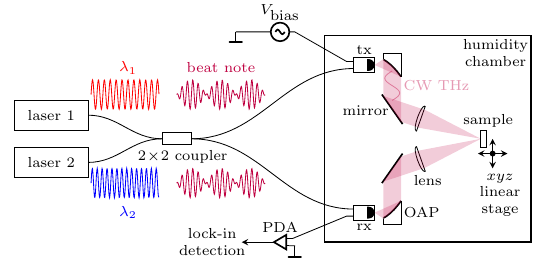}
\caption{Illustration of the experimental setup. Reproduced from Ref. \cite{Kristensen2024}, licensed under CC BY 4.0.}
\label{fig:1}
\end{figure}
The THz reflection spectra were acquired using continuous-wave (CW) THz frequency-domain spectroscopy (THz-FDS) over a frequency range of 0.09 to 1.19~THz. The measurements were performed using a TeraScan 1550 system (Toptica Photonics), employing an approximate incidence angle of 11$^\circ$. A schematic of the experimental setup is provided in Figure~\ref{fig:1}.
To ensure precise control over environmental conditions, the THz beam path was enclosed within a custom-designed humidity chamber. This chamber allowed for the adjustment of relative humidity (RH) levels between 5\% and 95\%, with a precision of $\pm$2\%, by purging the system with either dry or water vapor-saturated nitrogen.

For this study, five materials exhibiting spectral features within the operational bandwidth were selected; galactitol, L-tartaric acid (L-TA), 4-aminobenzoic acid (PABA), theophylline, and $\alpha$-lactose monohydrate. To prepare the samples, six pellets were fabricated for each compound at weight concentrations of 20\%, 50\%, and 80\%, using polyethylene (PE) powder as a binder. Additionally, two pure PE pellets were produced to serve as a control. PE was chosen for its flat spectral response in the THz range, ensuring minimal spectral interference \cite{Hua2010}.
The pellets were formed into 15$^\circ$ wedges to mitigate interference from multiple reflections within the sample. The sample holder featured a large aperture, significantly exceeding the THz beam diameter, to reduce diffraction effects and maintain consistent measurement conditions.
Each sample was analyzed across the full spectral range with a frequency step size of 80 MHz and an integration time of 3 ms per acquisition. To minimize spatial correlation effects, measurements were conducted at randomized positions across the sample surface.

Two distinct datasets were collected for training and validating the feature selection algorithms:
\begin{enumerate}[label=\Roman*]
    \item \emph{The training dataset} comprised 1920 spectra, obtained by measuring each sample 20 times under controlled RH conditions (10\%, 50\%, and 90\%), leading to 360 spectra per material (120 for pure PE).
    \item \emph{The test dataset} consisted of 2560 spectra, measured under ambient conditions with 80 acquisitions per sample, yielding 480 spectra per material (160 for pure PE).
\end{enumerate}
Since we seek to benchmark the performance of the algorithms to traditionally referenced data, a reference spectrum  was recorded using an aluminum mirror every 40 (20) measurements for the training (testing) data.
In total, the two datasets contained nearly 4500 spectra, with a training-to-testing ratio of 3:4. The full dataset is publicly accessible as part of the \emph{Database of frequency-domain terahertz reflection spectra for the DETRIS project} \cite{DETRISdb}. Further details on the experimental setup and data collection methodology can be found in Refs. \cite{Cielecki2021, Kristensen2024, DETRISdb}.

\subsection{Data Preprocessing}
The THz-FDS system employs a coherent detection scheme, which introduces phase oscillations in the recorded photocurrent $I_\mathrm{ph}(\nu)$ as the THz frequency $\nu$ is scanned. In this study, we extracted the instantaneous amplitude $A(\nu)$ and instantaneous phase $\phi(\nu)$ by applying the Hilbert transformation $\mathcal{H}$ to the oscillatory photocurrent, which is proportional to the THz electric field \cite{Vogt2017,Vogt2019}. Notably, the use of the Hilbert transform ensures that the spectral resolution remains equal to the optical frequency step size, independent of the THz optical path length.
The resulting complex-valued analytic signal,
\begin{equation*}
    I_\mathrm{a}(\nu)
    =
    I_\mathrm{ph}(\nu)+i\mathcal{H}\left\{I_\mathrm{ph}(\nu)\right\}
    =
    A(\nu)\exp\left[i\phi(\nu)\right],
\end{equation*}
was subsequently transformed into the time domain via a Fourier transform, allowing for the removal of unwanted reflections within the setup that could introduce Fabry-Pérot interference. This preprocessed signal was then inverse Fourier transformed back to the frequency domain \cite{Kong2018}. However, as excessive filtering could degrade the effective spectral resolution, only specific interference components were removed.

To focus on the spectral region relevant for material identification, the data was cropped to 0.4–1.05~THz, where the sample characteristics are most pronounced, while also optimizing the signal-to-noise ratio (SNR). Following our previous methodology \cite{Cielecki2021,Kristensen2024}, we utilized only the spectral amplitude $A(\nu)$ for further processing. However, it should be noted that, similar to THz time-domain spectroscopy (THz-TDS), material-specific information is also encoded in the spectral phase $\phi(\nu)$ \cite{Vogt2017,Vogt2019}.

The reflection coefficient was obtained by normalizing the sample spectrum against an appropriate reference spectrum:
\begin{equation}
    r(\nu)=A_\mathrm{sample}(\nu)/A_\mathrm{reference}(\nu).
    \label{eq1}
\end{equation}

Before calculating $r(\nu)$, the data was interpolated onto integer GHz frequencies to ensure accurate spectral deconvolution. Each spectrum consisted of 649 discrete frequency points. Throughout this work, we classify $A(\nu)$ as non-referenced spectral data and $r(\nu)$ as referenced spectral data to distinguish between raw and deconvoluted spectra.

\begin{figure}
\center
\includegraphics[width=\columnwidth]{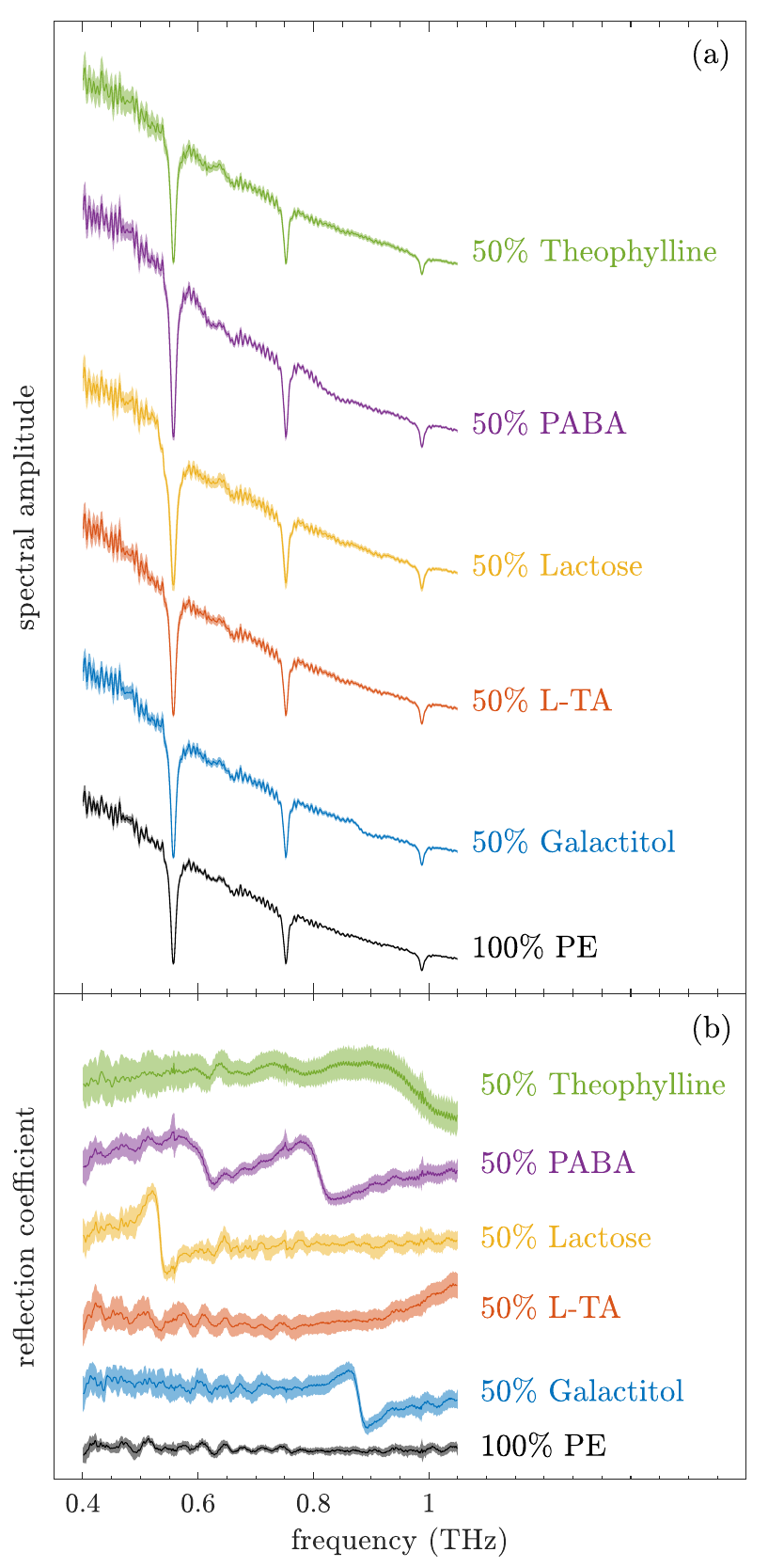}
\caption{
Non-referenced (a) and referenced (b) THz reflection spectra of the samples with 50\% of active material and pure PE measured under ambient conditions.
The dark colored center line of each curve is the mean of 160 measurements, while the light colored fill represents the standard deviation. The curves of each material are shifted vertically for a better readability.
Reproduced from Ref. \cite{Kristensen2024}, licensed under CC BY 4.0.
}
\label{fig:2}
\end{figure}
Figure~\ref{fig:2}(a) shows the non-referenced CW THz reflection spectra for samples containing 50\% active material and pure PE, acquired under ambient conditions. 
The absorption features of atmospheric water vapor are clearly visible around 0.55, 0.75, and 0.99~THz, while the characteristic spectral features of the materials themselves are less distinct.
The corresponding referenced spectra are shown in Figure~\ref{fig:2}(b). After deconvolution with reference spectra, the material-specific absorption characteristics  become more apparent. However, as the active material concentration decreases from 50\% to 20\%, the spectral differences between certain materials (e.g., theophylline, L-TA, and PE) become more difficult to resolve \cite{Cielecki2021}.
It is important to emphasize that both referenced and non-referenced spectra originate from the same raw data.

\subsection{Feature Selection Algorithms}
Feature selection is a key dimensionality reduction technique used to enhance the efficiency and accuracy of ML models. By identifying and retaining only the most informative variables – while discarding noisy or redundant ones – feature selection improves model generalization, reduces overfitting, and lowers computational demands 
\cite{Venkatesh2019,Li2017,Aggarwal2014,Guyon2003}. Unlike feature extraction methods such as PCA or LDA, which generate new features from combinations of original inputs, feature selection preserves the physical interpretability of the data by selecting a subset of the original features \cite{Aggarwal2014,Alpaydin14}.

In the context of THz reflection spectroscopy, each feature corresponds to a discrete frequency component in the measured spectrum.
Accordingly, we denote the data matrix by $X \in \mathbb{R}^{N \times M}$, where each row $n$ represents a THz spectrum $S_{\!n}(\nu)$ (i.e. an observation), and each column $m$ corresponds to a discrete frequency component $\nu_m$ (i.e. a variable). Moreover, each spectrum $S_{\!n}(\nu)$ is associated to a class $C_k\in\{C_1,C_2,\ldots,C_K\}$.

Feature selection techniques are commonly grouped into three categories based on how they interact with the classification model \cite{Guyon2003}:
\begin{itemize}
    \item \emph{Filter methods}, which rank features based on statistical criteria independent of the classifier.
    \item \emph{Embedded methods}, which perform feature selection intrinsically during the training of the classifier itself.
    \item \emph{Wrapper methods}, which evaluate subsets of features using classifier performance as a selection criterion.
\end{itemize}
This study compares representatives from each category – mRMR, LASSO regularization, and SFS – in combination with widely used classifiers to identify minimal, high-performance spectral subsets for reference-free THz material classification.

\subsubsection{Filter Method: mRMR}
Filter methods perform feature selection independently of the classification model, ranking features based on statistical properties such as correlation or mutual information \cite{Guyon2003}. This approach enables fast, classifier-agnostic feature selection but may neglect interactions between features and the model, occasionally leading to suboptimal classification performance. Common filter techniques include Relief, Information Gain,  and mRMR \cite{Aggarwal2014}.

Minimum Redundancy 
Maximum Relevance (mRMR) is a multivariate filter algorithm originally introduced by Peng \emph{et al.} \cite{Peng2003,Peng2005} for gene selection in microarray data, and has since been adopted in various domains \cite{Ramirez-Gallego2017}.
mRMR selects a feature subset $\mathcal{S} \subset \{\nu_1, \nu_2, \ldots, \nu_M\}$ that jointly maximizes relevance to the target class $C$ while minimizing redundancy among the selected features $\nu_m$. Both criteria are defined via the mutual information $I$, defined as:
\begin{equation}
I(x,y) = \sum_{i,j} p(x_i,y_j) \log\left(\frac{p(x_i,y_j)}{p(x_i)p(y_j)}\right),
\end{equation}
where $p(x_i)$ and $p(y_j)$ denote marginal probabilities and $p(x_i,y_j)$ is the joint distribution between two variables $x$ and $y$.

In this work, we used the Matlab implementation of mRMR based on Ref. \cite{Peng2003}, which ranks features using the mutual information quotient (MIQ):
\begin{equation}
\mathrm{MIQ}(\nu_m) = \frac{V(\nu_m)}{W(\nu_m)},
\end{equation}
where $V(\nu_m) = I(\nu_m, C_k)$ represents the relevance of feature $\nu_m$, and 
\begin{equation}
    W(\nu_m) = \frac{1}{\abs{\mathcal{S}}}\sum_{\nu \in \mathcal{S}} I(\nu_m,\nu)
\end{equation}
quantifies redundancy with respect to already selected features.
The mRMR algorithm was exclusively applied on the training set.

\subsubsection{Embedded Method: LASSO}
Embedded methods perform feature selection as part of the model training process \cite{Guyon2003}.
Popular embedded techniques include tree-based methods \cite{Guyon2003,Venkatesh2019} and regularized linear models, such as LASSO \cite{Tibshirani1996} and Elastic Net \cite{Zou2005}.

The Least Absolute Shrinkage and Selection Operator (LASSO) was introduced by Tibshirani in 1996 \cite{Tibshirani1996}, building on earlier work using $L_1$-regularization for inverse problems in geophysics \cite{Santosa1986}. LASSO has since become widely used for feature selection in generalized linear models, including logistic regression \cite{Friedman2010}. It imposes an $L_1$-norm penalty on the regression coefficients:
\begin{equation}
\mathrm{penalty}(\beta) = \lambda \sum_i \, \abs{\beta_i},
\end{equation}
where $\beta_i$ denotes the coefficient for feature $i$ and $\lambda$ is the regularization parameter. The $L_1$-penalty drives sparsity in the solution by shrinking some coefficients to zero, thereby performing feature selection \cite{James2013,Bishop06}.

We employed the \texttt{Glmnet} implementation in Matlab \cite{Friedman2010} to perform LASSO-regularized multinomial logistic regression. The model was trained using 71 values of $\lambda$ logarithmically spaced between $10^{-2}$ and 100. To ensure consistent feature selection across all class comparisons, we used the ‘grouped’ penalty option, which jointly retains or discards the coefficients associated with a given feature. The trained models were then evaluated on the test set. For each model, we recorded the classification accuracy and number of selected features to assess overall performance.

\subsubsection{Wrapper Method: SFS}
Wrapper methods perform feature selection by directly evaluating subsets of features based on the classification performance of a specific model \cite{Guyon2003}. Unlike filter methods, which rely solely on the statistical properties of individual features, wrappers account for feature interactions and their impact on the classifier. While typically yielding higher accuracy, this approach is computationally more intensive due to the need for repeated model training. Since the number of possible feature subsets scales exponentially with the number of features ($2^n$), exhaustive search is infeasible for large datasets, necessitating heuristic search strategies and appropriate stopping criteria.
Two commonly used greedy search strategies are \cite{Guyon2003}:
\begin{itemize}
\item \emph{Forward Selection:} Begins with an empty feature set and iteratively adds the feature that results in the greatest improvement in classification performance.
\item \emph{Backward Elimination:} Starts with all features and sequentially removes the least informative ones until a termination criterion is met.
\end{itemize}
More advanced techniques, such as Simulated Annealing, Branch-and-Bound, and Evolutionary algorithms (e.g., genetic algorithms and ant colony optimization),  offer more exhaustive exploration but at higher computational cost. Wrapper methods often outperform filter methods in terms of classification accuracy at the cost of increased complexity. \cite{Venkatesh2019}

In this study, we employed Sequential Forward Selection (SFS) combined with either a one-vs-one linear LR, a NB, or a SVM classifier (see below). 
The forward selection strategy was chosen based on performance trends observed in the other feature selection experiments (see Results section).
Starting from an empty set, features were added iteratively, selecting at each step the feature whose addition yielded the highest classification accuracy on the validation data.
A maximum of 20 features was allowed, serving as the stopping criterion for the SFS algorithm. 
To reduce computation time, all training and evaluation steps were executed using parallel processing.

\subsection{Classification Algorithms}
To evaluate the selected feature subsets, we employed three commonly used classification algorithms. Namely, linear Logistic Regression (LR), Naïve Bayes (NB), and Support Vector Machines (SVM). All classifiers were implemented using Matlab’s built-in functions, and where applicable, models were trained exclusively on the training set and evaluated on the independent test set.

\subsubsection{Linear Logistic Regression}
Linear LR is a binary classification model that assumes a linear relationship between the feature vector $S_{\!n}$ and the log-odds ratio of class membership \cite{Bishop06}:
\begin{equation}
    \log\left( \frac{p(C_1 | S_{\!n})}{p(C_2 | S_{\!n})} \right) = \beta_0 + S_{\!n}^\top \beta,
\end{equation}
where $p(C_k | S_{\!n})$ are the class-conditional probabilities.
Applying the logistic function to this linear predictor yields:
\begin{align}
    p(C_1 | S_{\!n}) &= \frac{1}{1 + \exp\left[ -(\beta_0 + S_{\!n}^\top \beta) \right]}, \\
    p(C_2 | S_{\!n}) &= 1 - p(C_1 | S_{\!n}),
\end{align}
where $\beta_0$ is the intercept and $\beta \in \mathbb{R}^M$ is the vector of regression coefficients. 
Hence, this model induces linear decision boundaries between the classes.
Because LR is inherently binary, several extensions have been proposed for multiclass classification tasks. A common approach is \emph{problem binarization} \cite{Galar2011,Lei2019}, which decomposes the $K$-class problem into multiple binary ones. Two widely used strategies are \cite{Galar2011,Lorena2009}:
\begin{itemize}
    \item \emph{One-vs-one}: Constructs a binary classifier for each pair of classes, resulting in $\binom{K}{2}$ models. Final classification is based on majority voting among all pairwise classifiers.
    \item \emph{One-vs-all}: Builds $K$ classifiers, each trained to distinguish one class from the remaining $K-1$. The class with the highest predicted probability is selected.
\end{itemize}
An alternative is multinomial LR, which extends the model to $K$ classes by expressing $K - 1$ log-odds ratios relative to a reference class $C_K$ \cite{Lei2019}:
\begin{equation}
    \log\left( \frac{p(C_k | S_{\!n})}{p(C_K | S_{\!n})} \right) = \beta_{0k} + S_{\!n}^\top \beta_k, \qquad k = 1, \dots, K-1.
\end{equation}
In this work, we applied one-vs-one linear LR models with ridge regularization to mitigate overfitting
For each feature subset, 21 models were trained using regularization parameters logarithmically spaced between $10^{-4}$ and $10^{-1}$. 
The best-performing model on the test set determined the accuracy associated with the feature subset.

\subsubsection{Naïve Bayes}
The NB classifier is based on Bayes’ theorem, computing the posterior probability of an observation $S_{\!n}(\nu)$ belonging to class $C_k$ as:
\begin{align}
    &
	p(C_k | S_{\!n}) = \frac{p(S_{\!n}| C_k) p(C_k)}{p(S_{\!n})}
    \\[.2cm]
    &
	\quad
    \text{with}
    \quad
	p(S_{\!n}) = \sum_j p(S_{\!n} | C_k) p(C_k),
    \notag
\end{align}
where $p(C_k)$ is the prior probability of class $C_k$, and $p(S_{\!n} | C_k)$ is the likelihood of observing $S_{\!n}$ given class $C_k$ \cite{Alpaydin14}. The observation is assigned to the class with the highest posterior probability.
In this study, we assumed the class-conditional likelihoods to follow multivariate Gaussian distributions with class-specific means and covariance matrices estimated from the training data.

\subsubsection{Support Vector Machine (SVM)}
Support vector machines classify observations by constructing a hyperplane 
\begin{equation}
    H:\quad \vec w\cdot S_{\!n} + b = 0
\end{equation}
that maximizes the margin between observations from different classes \cite{Cortes1995,Burges1998}.
Here, $\vec w$ is a normal vector of the hyperplane and $b$ is the bias term. 
When an optimal hyperplane is determined, it serves as the decision boundary, and new observations are classified according to the side of the boundary on which they lie, i.e. class labels reduce to $y_n=\pm1$.
Thus, SVM is fundamentally binary but can be extended to a multi-class modality through one-vs-one or one-vs-all strategies, like linear LR.
For linearly separable data, the optimal hyperplane is found by solving:
\begin{equation}
\underset{y_n(\vec{w} \cdot S_{\!n} + b) \geq 1}{\arg\min} \left( \frac{1}{2} |\vec{w}|^2\right).
\end{equation}
The notation arg min refers to the set of input values that minimize the given function.

In practice, many real-world datasets are not linearly separable. To improve generalization, a soft-margin approach is typically used, allowing some margin violations via slack variables $\xi_n\geq 0$:
\begin{equation}
\underset{y_n(\vec{w} \cdot S_{\!n} + b) \geq 1}{\arg\min} \left( \frac{1}{2} |\vec{w}|^2 + \gamma \sum_n \xi_n \right),
\end{equation}
where $\gamma > 0$ controls the trade-off between maximizing margin and minimizing classification error.

In this study, we employed soft-margin SVMs with a linear kernel and one-vs-one strategy.
In all cases, the regularization parameter $\gamma$ was optimized using 10-fold cross-validation during, applied during model fitting (i.e. not during feature selection).

\subsection{Implementation Details}
All algorithms were implemented in MATLAB (R2022b). Computations were performed on an Apple MacBook Pro equipped with an 2.8 GHz Quad-Core Intel Core i7 CPU and 16 GB RAM, running MacOs Ventura 13.7.4.

\section{Results and Discussion}
In the following, we evaluate the performance of the feature selection methods in combination with the classification algorithms. 
Since classification accuracy is used as the performance criterion across all methods, classifier behavior and feature selection performance are inherently coupled.
As such, we present and discuss these results jointly, rather than isolating classifier performance from the selection process.

\subsection{Filter and Embedded Methods: mRMR and LASSO}
\begin{figure}
\center
\includegraphics[width=\columnwidth]{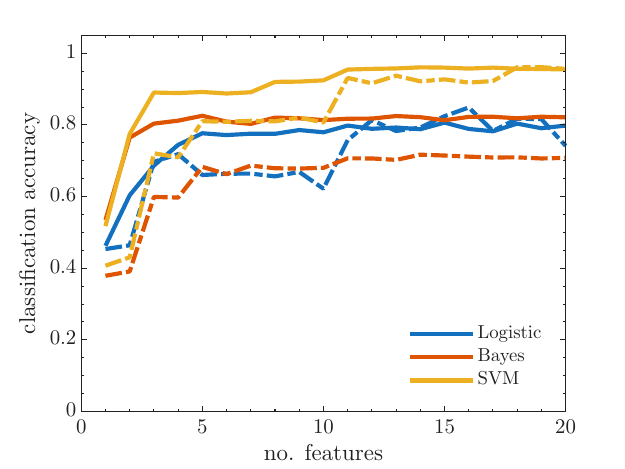}
\caption{
Classification accuracy as a function of the number of selected features by the mRMR algorithm, evaluated on both referenced (solid lines) and non-referenced (dashed lines) THz spectra. 
Results are shown for logistic regression (blue), Naïve Bayes (red), and support vector machine (SVM, yellow) classifiers. 
}
\label{fig:3}
\end{figure}

Figure~\ref{fig:3} presents the performance of the mRMR filter algorithm evaluated across the three classifiers on both referenced (solid lines) and non-referenced (dashed lines) THz spectra.
On referenced data, all classifiers demonstrated rapid initial gains, plateauing after approximately five features, indicating that most relevant discriminative information was captured within this compact subset.
Over 20 features, LR improved from 46.3\% to 79.8\% accuracy, and NB from 53.4\% to 82.1\%.
Noteworthy for SVM, accuracy increased from 51.8\% with a single feature to 89.1\% by the third, reaching a peak of 96.1\% with 14 features. The performance plateaued beyond this point with minimal degradation (above 95\% through feature 20), indicating strong model stability.
On non-referenced spectra, performance remained competitive or even exceeded the referenced results.
Particularly for SVM, the accuracy improve gradually, reaching 81\% by the fifth feature, and remained consistently above 91\% after the tenth feature, with a maximum of 96.1\% achieved at feature 18.
LR improved from 45.4\% to 79.8\%, closely matching its referenced counterpart, while NB rose from 37.9\% to 70.8\% over the same feature span. 
These findings demonstrate that mRMR consistently identifies a sparse yet highly informative set of frequencies, enabling robust classification with as few as 10–20 inputs.
They further affirm that accurate material identification is feasible even under non-referenced conditions, particularly when using classifiers such as SVM that can exploit complex patterns within low-dimensional feature spaces.
Overall, the mRMR-selected frequencies generalize well under cross-validated SVM training, supporting reliable classification without referencing while using roughly 2\% of the complete set of spectral components.

\begin{figure}
\center
\includegraphics[width=\columnwidth]{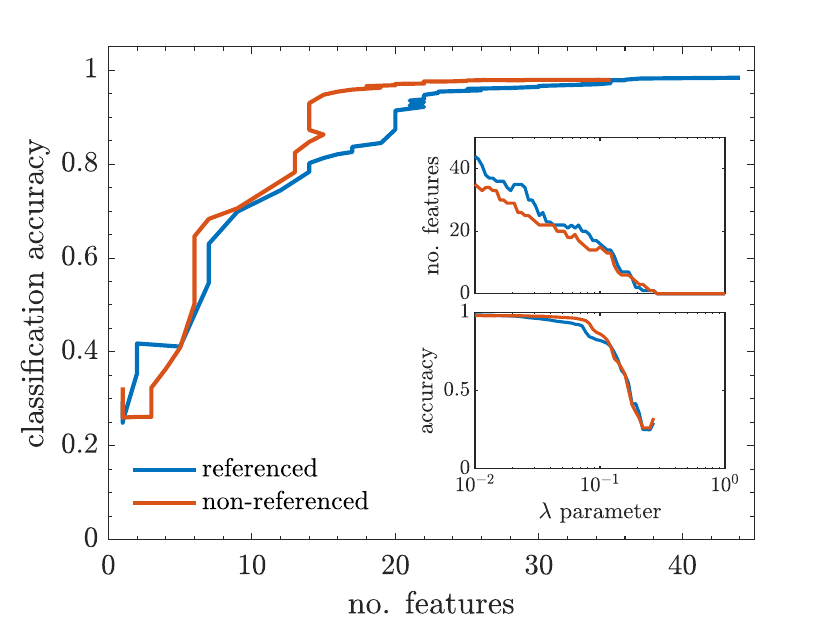}
\caption{
Classification accuracy as a function of the number of features selected by the LASSO algorithm. The blue curve corresponds to referenced THz spectra, and the red curve to non-referenced THz spectra. The inset shows the cross-validated selection of the optimal regularization parameter $\lambda$.
}
\label{fig:4}
\end{figure}

For the embedded LASSO method, the classification accuracy improved progressively with the number of selected features for both referenced and non-referenced spectra (Figure~\ref{fig:4}).
In the referenced case, the accuracy increased rapidly beyond five features, reaching above 90\% with about 20 features and converging toward a maximum of 98.4\% with 44 features. 
The non-referenced spectra followed a similar trend, exceeding 93\% after 14 features and saturating near 98.0\% with 29 features.
The insets in Figure~\ref{fig:4} show the evolution of accuracy and the number of selected features as a function of the regularization parameter $\lambda$. 
As $\lambda$ decreases, more features are selected, and accuracy rises until reaching diminishing returns. For the referenced spectra, accuracy plateaus around $\lambda=$0.010--0.0139, corresponding to 37--44 features; for non-referenced spectra, a similar plateau occurs around $\lambda=$0.010--0.0181 with 29--35 features. 
These plots illustrate how the optimal $\lambda$ balances dimensionality reduction with classification performance. Overall, the LASSO approach achieved high accuracy with relatively few spectral features, demonstrating its effectiveness in reducing dimensionality while maintaining strong classification performance.

Unlike filter and wrapper methods, LASSO does not directly control the number of selected features. Instead, selection is governed by the regularization parameter $\lambda$, which limits the sum of the absolute values of the regression coefficients.
This can produce irregular classification curves as seen in Figure~\ref{fig:4}.
That is, for some $\lambda$ values, the algorithm selects the same number of features, while other feature counts are skipped. Decreasing $\lambda$ may occasionally revisit smaller feature subsets but with higher coefficient values. These behaviors arise from LASSO’s dual role as a feature selector and regularizer, where coefficients are shrunk toward zero, effectively removing some features, while non-zero coefficients remain constrained by $\lambda$. Consequently, the classification curves increase less steeply in the early phase compared to filter-based methods, and the number of selected features does not vary monotonically with $\lambda$.

Compared to the filter-based mRMR approach, the embedded LASSO method required a larger number of features to reach its maximum accuracy. mRMR achieved near-optimal performance within a compact subset of 10–20 frequencies, with accuracies saturating around 95–96\% for both referenced and non-referenced data. By contrast, LASSO exhibited a more gradual accuracy gain, surpassing 90\% only after approximately 18–20 features and converging toward its maximum at around 25–35 features, with peak accuracies of 98.4\% (referenced) and 98.0\% (non-referenced). Thus, while mRMR offered highly efficient dimensionality reduction with minimal feature sets, LASSO provided slightly higher ultimate accuracy at the expense of a larger feature subset. These complementary behaviors highlight a trade-off between compactness and peak classification performance.

\subsection{Wrapper Method: Sequential Forward Selection}
\begin{figure}
\center
\includegraphics[width=\columnwidth]{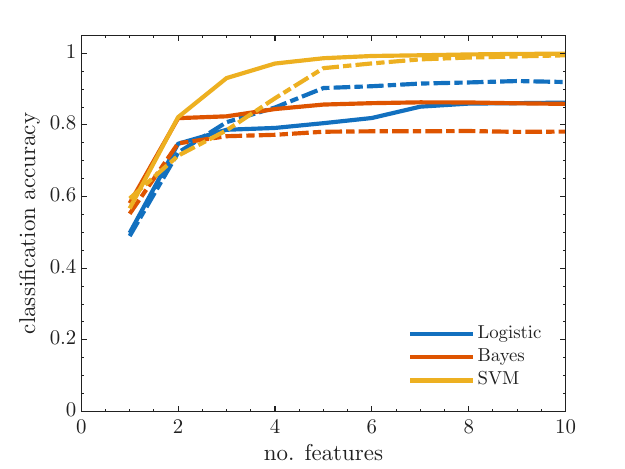}
\caption{
Classification accuracy as a function of the number of selected features by the SFS algorithm, evaluated on both referenced (solid lines) and non-referenced (dashed lines) THz spectra. 
Results are shown for logistic regression (blue), Naïve Bayes (red), and support vector machine (SVM, yellow) classifiers.
}
\label{fig:5}
\end{figure}

The performance of the wrapper-based sequential forward selection (SFS) method is presented in Figure~\ref{fig:5} for the LR (blue), NB (red), and SVM (yellow) classifiers.
A consistent trend is seen across all classifiers and both data types.
The accuracy increases sharply within the first two to five selected features, after which it reaches a plateau.
For the referenced spectra (solid lines), all three classifiers exhibit steady gains in accuracy as additional features are included, with SVM achieving the highest accuracy of 99.9\% at ten features, followed by LR at 86.2\% and NB at 85.9\%.
SFS also yields superior performance for the non-referenced spectra (dashed lines), particularly for SVM, which reaches nearly perfect classification (99.5\%) with ten features.
Overall, the non-referenced SVM accuracy approaches that of the referenced case with just a few additional inputs.
Given that the original spectra consist of 649 interpolated components spaced at 1-GHz intervals, this result indicates that only 1\% of the total spectral data is sufficient for near-perfect classification, underscoring the effectiveness of the feature selection approach.
LR and NB also improve markedly, achieving 92.0\% and 78.1\%  accuracy at ten features.

Interestingly, the LR and NB classification performance is lower for the referenced data. This result may initially appear counterintuitive, as referencing is typically intended to remove systematic effects. However, referencing also removes critical variability (such as contributions from the system response) that the feature selection algorithm exploits to discriminate between classes. Thus, while referencing may simplify the spectra, it inadvertently reduces the information content relevant for classification.

NB consistently performs the worst on the non-referenced data. This is expected given that NB assumes the input features are independent and identically distributed, which is clearly violated in the presence of a correlated system response. In contrast, NB performs better on the referenced data, where such correlations have been minimized. LR, although also a probabilistic method, does not require feature independence. As a result, it achieves higher accuracy than NB but remains susceptible to overfitting in the presence of highly correlated features. SVM on the other hand, which makes no assumptions about feature distributions and incorporates a soft-margin regularization strategy, consistently outperforms both LR and NB across all scenarios.

Compared to mRMR and LASSO, SFS generally achieves higher accuracies, especially in the non-referenced case, and demonstrates faster convergence to high performance with relatively few features.
Despite these high accuracies, the computational cost of training the wrapper-based SFS is significant. For LR, SFS required 3--5 hours per dataset, NB took 30 minutes, whereas SVM ranged from 7 hours to 4.5 days.
By comparison, filter-based mRMR required 6–35 minutes, while the embedded LASSO method was extremely fast ($\sim$2 seconds) but with lower overall accuracy. 
This emphasizes that wrapper-based feature selection is particularly well suited for THz spectral data when maximizing classification accuracy is prioritized over computational cost in the training phase.

\subsection{Discriminative Spectral Components Identified by SFS}
\begin{figure}
\center
\includegraphics[width=\columnwidth]{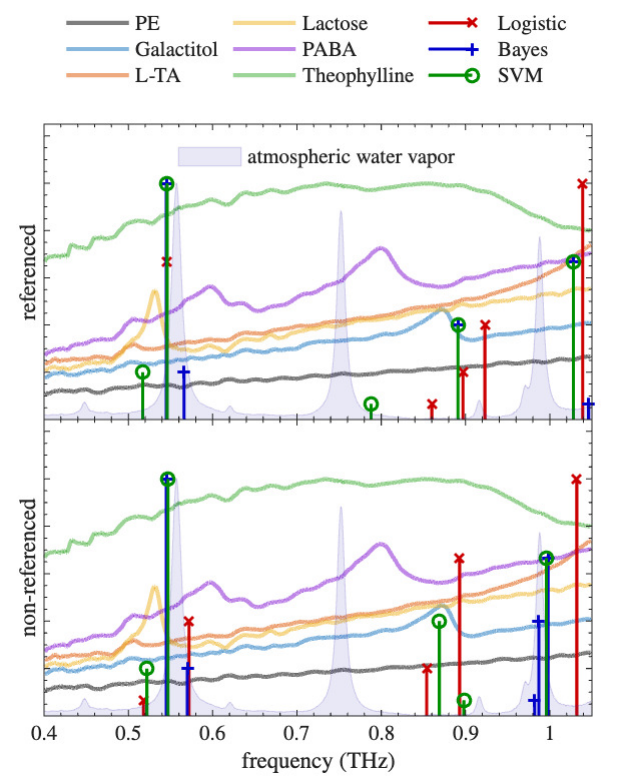}
\caption{
Frequency-dependent discriminative power of the five most dominant features selected by the SFS algorithm for the different classifiers and scenarios (referenced and non-referenced data), together with the absorption coefficients of the materials (normalized to that of theophylline).
}
\label{fig:6}
\end{figure}

To further analyze the behavior of the SFS algorithm, we plot the five most discriminative features (vertical bars) identified for each classifier in Figure~\ref{fig:6}, overlaid on the absorption coefficients of the six materials to enable direct physical interpretation of the selected frequencies.
The height of each bar reflects the feature's relative contribution to the classifier's performance, revealing how the SFS algorithm prioritizes spectral components.
A strong correspondence is observed between the selected frequencies and the absorption bands of the materials.
Several features were selected around 0.53–0.57, 0.86–0.90, and 1.0–1.05~THz, aligning with the spectral characteristics of lactose (0.53~THz), PABA (0.6 and 0.8~THz), galactitol (0.87~THz), theophylline (0.9~THz), and L-TA ($\sim$1.1~THz).
Although lactose, galactitol, and PABA exhibit strong absorption peaks at 0.53, 0.87, and 0.8~THz respectively, SFS did not generally select these exact frequencies, likely because their contribution was redundant with previously chosen features or did not provide sufficient incremental discriminative power.

The near-perfect performance of SVM on the non-referenced data is underpinned by the balanced feature distribution spanning both the low-frequency region around 0.52–0.55~THz and high-frequency range from 0.86~THz to 1.0~THz.
Meanwhile, the poor performance of the NB classifier for both referenced and non-referenced data can be explained by the clustering of multiple features in certain spectral bands (e.g., around 1~THz), leading to a worse generalization among the materials.

Overall, this correlation between SFS-selected frequencies and material absorption characteristics verifies the physical interpretability of the classification approach, confirming that the algorithm consistently identifies physically meaningful features independent of the referencing condition and that the discriminative power arises from genuine spectroscopic contrasts between materials.
Additionally, the water vapor absorption bands are largely omitted, most likely due to the variability of the spectral amplitude across measurements under different humidity conditions.

\section{Conclusion}
In this study, we have demonstrated that feature selection algorithms can enable accurate, reference-free classification of materials using sparse THz reflection spectroscopy. 
By evaluating filter-based (mRMR), embedded (LASSO), and wrapper-based (Sequential Forward Selection) strategies in combination with logistic regression, Naïve Bayes, and support vector machine classifiers, we show that high classification accuracy can be achieved even on non-referenced spectra.

Particularly, SFS facilitated near-perfect (99.5\%) classification accuracy for SVM on non-referenced spectra with only 1\% of the full spectral data (i.e. 10 features). An analysis of the SFS features revealed a clear correspondence between the selected frequencies and material absorption bands. A balanced feature distribution across these absorption bands consolidates superior performance, and vice versa, an inferior performance is linked to clustering of the selected features. Finally, these observations confirm that the selected frequencies are physically meaningful, reflecting genuine spectroscopic contrasts, while water vapor absorption bands are disregarded.

Our results show that carefully selected sparse-frequency subsets are sufficient for robust material discrimination, reducing reliance on broadband sources and reference measurements. 
While our work \cite{Kristensen2025_IRMMWTHz} provides a data-driven foundation, two contemporaneous studies \cite{Schwenson2025,Schwenson2025_IRMMWTHz} address the hardware implementation side, independently validating the feasibility of high-speed sparse-frequency THz sensing.
First, they implemented sparse THz frequency-domain sensing by selecting eight discrete frequencies across a 1.5-THz range, achieving a 90~dB peak dynamic range  in 800~ms with kilohertz measurement rates \cite{Schwenson2025}.
Subsequently, they demonstrated a photonic integrated circuit (PIC) enabling CW THz generation and coherent detection at kilohertz speeds \cite{Schwenson2025_IRMMWTHz}.
Together, these hardware advances and our data-driven results chart a clear path toward fast, compact THz sensors optimized for real-world applications such as security screening, non-destructive testing, and environmental monitoring.
Future work will focus on implementing these insights in hardware systems operating at the identified frequencies.

\section*{Funding} This work was supported by the Innovation Fund Denmark Grand Solutions program (grant no. IFD-7076-00017B).

\section*{Disclosures} The authors declare no conflicts of interest.

\section*{Data availability} Data underlying the results presented in this paper are publicly available at DOI: \href{https://doi.org/10.5281/zenodo.5079558}{10.5281/zenodo.5079558}.

\section*{CRediT Authorship Contribution Statement}
\textbf{Mathias Hedegaard Kristensen:} Conceptualization, Investigation, Methodology, Validation, Formal Analysis, Software, Visualization, Writing – original draft, Writing – review \& editing.
\textbf{Paweł Piotr Cielecki:} Conceptualization, Investigation, Methodology, Writing – review \& editing.
\textbf{Esben Skovsen:} Conceptualization, Funding Acquisition, Supervision, Writing – review \& editing.

 \bibliographystyle{elsarticle-num} 
 \bibliography{featureselection}





% \end{thebibliography}
\end{document}